\documentclass[prb,twocolumn,showpacs,preprintnumbers,amsmath,amssymb]{revtex4}

\usepackage{graphicx}
\usepackage{dcolumn}
\usepackage{bm}
\usepackage[]{ams,amsmath,amssymb,epsfig}

\begin{document}
\title{Single-particle states in spherical Si/SiO$_2$ quantum dots}

\author{A.S. Moskalenko}
 \altaffiliation[Also at\;]{Ioffe Physico-Technical Institute of RAS, 26
Polytechnicheskaya, 194021 St.~Petersburg, Russia}
\email{moskalen@mpi-halle.de}

\author{J. Berakdar}

\affiliation{Max-Planck-Institut f\"{u}r Mikrostrukturphysik,
06120 Halle \& Institut f\"ur Physik, Martin-Luther-Universit\"at
 Halle-Wittenberg, Nanotechnikum-Weinberg, Heinrich-Damerow-Str.~4, 06120 Halle, Germany }
\author{A.A. Prokofiev}
\author{I.N. Yassievich}
\affiliation{Ioffe Physico-Technical Institute of RAS, 26
Polytechnicheskaya, 194021 St.~Petersburg, Russia}

\date{\today}

\begin{abstract}
We calculate ground and excited electron and hole levels in
spherical Si quantum dots inside SiO$_2$ in a multiband effective
mass approximation. Luttinger Hamiltonian is used for holes and
the strong anisotropy of the conduction electron effective mass in
Si is taken into account. As boundary conditions for electron and
hole wave functions we use continuity of the wave functions and
the velocity density at the boundary of the quantum dots.
\end{abstract}



\maketitle


\section{Introduction}
The study of  materials composed of Si nanocrystals dispersed in
SiO$_{2}$ matrix is an issue of high importance  for various
optoelectronic applications \cite{Pavesi,Polman_Nature}.  In
particular, the knowledge of the energy spectrum of carriers
confined in the nanocrystals and their wave functions is crucial
for the understanding of   electronic processes. Numerous
theoretical works have been developed for   the evaluations of the
ground-state electron-hole pair energy of Si nanocrystals
\cite{Proot,Martin,Delley1993,Wang,Wolkin,Garoufalis2006}. On the
other hand,  the problem concerning  the energy-level positions
and the corresponding eigenfunctions of the excited carrier states
is much less studied \cite{Niquet,Burdov}. However, shortly after
the generation of an electron-hole pair in a nanocrystal a number
of  important non-equilibrium processes involving these ''hot''
carriers may take place \cite{Nacho_PRL} necessitating thus the
knowledge of the excited states.

This current paper is devoted to the study of the ground and the excited
electron and hole levels within  the framework of a multiband effective mass
approximation. The finite energy barriers at the Si/SiO$_2$ boundary are
explicitly accounted for.
 For the description of the electron and hole states of the carriers
confined in Si nanocrystals we utilize  the envelope function
approximation taking into account the elliptic symmetry of the
bottom of the conduction band and the complex structure of the top
of the valence band in Si. The finite energy barriers at the
boundary between Si and SiO$_2$ are treated by employing  the
Bastard boundary conditions \cite{Cardona_book,Ivchenko}. This is
 mainly the difference of our method comparing to  earlier
calculations based on the effective mass approximation \cite{Niquet} that
failed to describe properly the optical properties of small nanocrystals. An
advantage of our method in comparison with  {\it ab-initio} methods based on
the density functional theory \cite{Delley1993,Garoufalis2006} is that without
a considerable numerical effort we can  calculate not only the ground state but
also excited states of the confined carriers. Furthermore, our theory can be
applied to a broad range of nanocrystal sizes. We are not limited to
nanocrystals with a small number of atoms. The other point is that the
calculations of various excitation and de-excitation processes under
participation of the confined carriers using the derived  wave functions are
transparent and allow for an insight into the underlying physics.

\section{Electron states}
The conduction band of bulk Si has six equivalent minima in the
first Brillouin zone at points $\pm\vec{k}_{0,z}=(0,0,\pm
0.85)k_X$, $\pm\vec{k}_{0,y}=(0,\pm 0.85,0)k_X$, and $\pm
\vec{k}_{0,x}=(\pm 0.85,0,0)k_X$, where $k_X=2\pi/a$ and
$a=0.543$~nm is the lattice constant of~Si~\cite{gammas}. The
minima are situated in the neighborhood of the six $X$-points
(there are three non-equivalent $X$-points). The conduction band
is doubly degenerate at each of the $X$-points, which is a
consequence of the fact that Si lattice has two atoms in the
elementary unit cell and the origin can be chosen at the center of
any of them. Assuming the Bloch amplitudes not changing in the
neighborhood of the $X$-point one can write the wave function of
one of the six equivalent ground states of electrons in the
nanocrystal as
\begin{equation}\label{Eq: electron_wave_function}
    \psi^e_\nu=\xi^e(\vec{r})u_{c\nu} {\rm e}^{i\vec{k}_{0,\nu}\vec{r}}
    \qquad
    (\nu=\pm x,\pm y,\pm z),
\end{equation}
where $u_{c\nu}$ is one of two Bloch amplitudes of bulk electron at $X$-point
in the Brillouin zone, which corresponds to the lower conduction band at the
$\vec{k}_{0,\nu}$ point. Note, that $u_{c\nu}$ yields zero overlap integral
with the Bloch amplitudes of the top of the valence band \cite{si-bands}. The
overlap integral of the Bloch amplitudes of the top of the valence band with
the second Bloch amplitude $u_{s\nu}=u_{c\nu}{\rm
e}^{-i2\vec{k}_{X,\nu}\vec{r}}$ corresponding to the upper conduction band at
$\vec{k}_{0,\nu}$ is not equal to zero.
The envelope wave function $\xi^e$ in Eq. \eqref{Eq:
electron_wave_function} inside the Si quantum dot satisfies the
following equation:
\begin{equation}\label{Eq:Shr_full}
  \begin{split}
    \frac{\hbar^2}{2m_{\|}}\frac{\partial^2}{\partial z^2}\xi^e(x,y,z)+
    \frac{\hbar^2}{2m_{\bot}}\left(\frac{\partial^2}{\partial x^2}+
     \frac{\partial^2}{\partial
     y^2}\right)\xi^e(x,y,z)&\\
     +E\xi^e(x,y,z)=0&,
   \end{split}
\end{equation}
where $m_\|=0.916 m_0, m_\bot=0.19 m_0$ with $m_0$ being the free
electron mass. The rigorous formulation of the boundary conditions
for the boundary between Si and SiO$_2$ in the framework of the
envelope function method is not a trivial task and generally has
to be investigated in comparison with experiment and numerical
methods. The Bastard type boundary conditions imply that $\xi$ and
$\hat{\vec{v}}\xi$ are continuous across the boundary, where
$\hat{\vec{v}}=\frac{1}{i\hbar}[\vec{r},\hat{H}]$ is the velocity
operator, $\xi$ is the envelope wave function and $\hat{H}$ is the
corresponding Hamiltonian \cite{Cardona_book,Ivchenko}. Here we
assume that the spectrum of electronic states outside the
nanocrystal in SiO$_2$ is isotropic and determined by a single
electron effective mass, which is equal to the free electron mass
$m_0$.
%
Then outside
the quantum dot we have
\begin{equation}\label{Eq:Shr_full_outside}
  \begin{split}
   \frac{\hbar^2}{2m_0}\left(\frac{\partial^2}{\partial x^2}+
    \frac{\partial^2}{\partial
    y^2}+\frac{\partial^2}{\partial
    z^2}\right)\xi^e(x,y,z)&\\
    +(E-U_e)\xi^e(x,y,z)=0&,
  \end{split}
\end{equation}
where $U_e$ is the energy barrier for electrons. According to
Ref.~\onlinecite{Burdov} we have $U_e=3.2$~eV. The boundary
conditions result in the following equations
\begin{equation}\label{Eq:boundary_conditon_electrons1}
    \xi^e\big|_{r=R_{\rm nc}^-}
    =\xi^e\big|_{r=R_{\rm nc}^+}\; ,
\end{equation}
\begin{equation}
    \left[\frac{1}{m_\bot}\frac{\rho}{R_{\rm nc}}
    \frac{\partial \xi^e}{\partial \rho}
    +
    \frac{1}{m_\|}\frac{z}{R_{\rm nc}}
    \frac{\partial \xi^e}{\partial z}\right]_{r=R_{\rm nc}^-}\hspace{-0.3cm}=\frac{1}{m_0}
    \frac{\partial \xi^e}{\partial r}\bigg|_{r=R_{\rm nc}^+}\; , \label{Eq:boundary_conditon_electrons3}
\end{equation}
which are most conveniently written in the cylindrical coordinate
system $(z,\rho,\phi)$.

\begin{figure}[t]
  \centering
  \epsfig{file=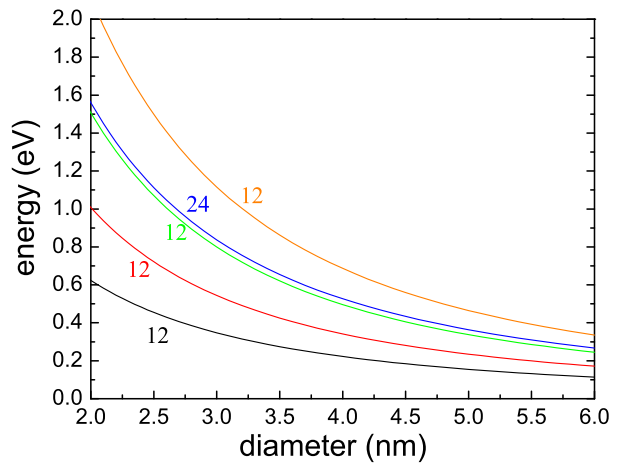,width=8.0cm}
  \caption{(Color online) Dependence of positions of the
electron energy levels above the bottom of the conduction band of
bulk-Si on the quantum dot diameter. The numbers near the lines
indicate the total degeneracy (including the spin degeneracy) of
the corresponding levels.}\label{Fig:levels_eh}
\end{figure}

\begin{figure*}[t]
  \centering
  \epsfig{file=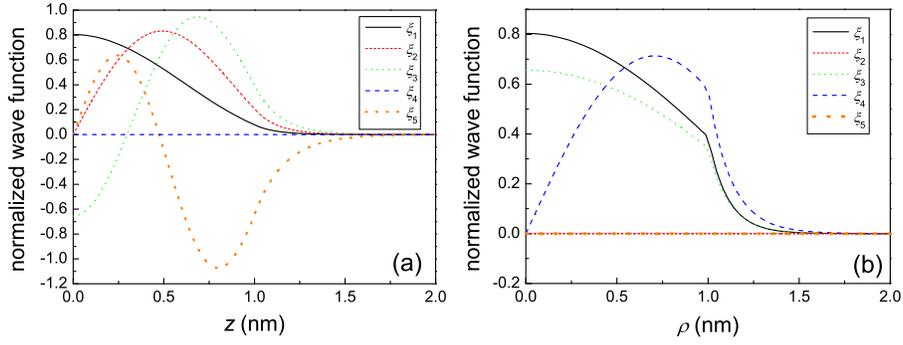,width=12.0cm}
  \caption{(Color online) Electron wave functions in dependence on the
  cylindrical coordinates $z$ (a) and $\rho$ (b)
  for five lowest electron levels of a quantum dot with diameter of
  2~nm. The wave function $\xi^e_4$ of the fourth level with  magnetic number $m=\pm1$ has
  also an angular dependence ${\rm e}^{im\phi}$.}\label{Fig:electron_functions}
\end{figure*}

Equations \eqref{Eq:Shr_full} and \eqref{Eq:Shr_full_outside} with
the boundary conditions \eqref{Eq:boundary_conditon_electrons1}
and \eqref{Eq:boundary_conditon_electrons3} have been solved
numerically after separating the trivial angular part
$\frac{1}{\sqrt{2\pi}}\exp(im\phi)$ ($m=0,\pm1,\pm2,...$) of the
wave functions. We have got the electron energy levels and the
corresponding envelope wave functions. The dependence of positions
of the several lowest energy levels on the quantum dot diameter is
depicted in Fig.~\ref{Fig:levels_eh}. The dependence of the
electron envelope wave functions on the distance from the center
of the quantum dot  is shown in Fig.~\ref{Fig:electron_functions}
for $d=2$~nm (such a small diameter is chosen for demonstration
reasons in order to resolve better the tunnelling tails of the
envelope wave functions). We have also compared the positions of
the electron levels and their degeneracies with the existing data
of Ref.~\onlinecite{Niquet} calculated by the empirical
tight-binding method for quantum dots with diameter $d=7.61$~nm.
We have found that apart from small level splittings due to the
valley-orbit interaction neglected in our model we get the same
sequence of levels. The levels are, however, shifted towards the
lower energies. The reason why we have smaller energies is that
this tight-binding model used truncation of Si nanocrystals by H
atoms. This procedure is known to give higher energies and greatly
overestimate the optical band gap when compared with experiments
on Si/SiO$_2$ nanocrystals and recent ab-initio TD-DFT
calculations \cite{Garoufalis2006} as well as with our model.


\section{Hole states}
For description of the valence band structure in Si we use a
generalization of the Luttinger Hamiltonian \cite{book_IN} in the
limit of a vanishing spin-orbit coupling, which is justified for
Si, i.e. we write the Hamiltonian $H$ in the form
\begin{equation}\label{Eq:Hamiltonian_valence}
 \hat{H}=(A+2B)\hat{p}^2-3B(\hat{\vec{p}} \cdot
 \hat{\vec{J}})^2,
\end{equation}
where $\hat{\vec{p}}$ is the momentum operator and $\hat{\vec{J}}$
is the unitary angular momentum operator acting in the space of
Bloch amplitudes. Furthermore, we introduced
\begin{equation}\label{Eq:AB}
A=-\frac{1}{4}\frac{m_h+m_l}{m_h m_l},\ \
B=-\frac{1}{4}\frac{m_h-m_l}{m_h m_l},
\end{equation}
\begin{equation}\label{Eq:m_h_m_l_gamma}
m_h=\frac{m_0}{\gamma_1-2\gamma}, \ \
m_l=\frac{m_0}{\gamma_1+2\gamma}, \ \ \
\gamma=\frac{1}{5}(3\gamma_3+2\gamma_2).
\end{equation}
Values of the constants $\gamma_1$, $\gamma_2$, and $\gamma_3$ for
Si are 4.22, 0.53, and 1.38, respectively \cite{gammas}.
The basis of the Bloch amplitudes space can be chosen in the form
of spherical components \cite{Edmonds} $u_0=Z$,
$u_\pm=\mp\sqrt{1/2}(X\pm iY)$ of the corresponding functions
$X=yz$, $Y=xz$, and $Z=xy$, of the representation $\Gamma_{25'}$
\cite{Cardona_book,si-bands}.

In the bulk Si this model leads to two types of states
corresponding to a doubly degenerate (in absence of spin-dependent
interactions) heavy hole band having mass $m_h$ and a
non-degenerate light hole band having mass $2m_hm_l/(3m_h-m_l)$.
The quantum confinement gives rise to mixing of the states.
Eigenfunctions of the Hamiltonian \eqref{Eq:Hamiltonian_valence}
can be found as eigenfunctions $\psi_{FM}$ of the square
$\hat{F}^2$ of the full angular momentum operator
$\hat{\vec{F}}=\hat{\vec{L}}+\hat{\vec{J}}$
($\hat{\vec{L}}=-i\vec{r}\times\partial_{\vec{r}}$) and its
projection $\hat{F}_z$ onto the axis $z$ \cite{Baldereschi}.
Eigenvalues of $\hat{F}^2$ and $\hat{F}_z$ are $F(F+1)$ and $M$,
respectively, where $F=0,1,2,...$ and  $M$ can be any integer
number having absolute value not larger than $F$. For a spherical
quantum dot there are three  types of states, namely
\begin{equation}\label{psi_S}
  \begin{split}
     \psi^{F-1,F+1}_{FM}(r,\theta,\phi)=&R^{F-1}_{F}(r){\bf Y}^{F-1}_{FM}(\theta,\phi)\\
    &+R^{F+1}_F(r){\bf Y}^{F+1}_{FM}(\theta,\phi),
   \end{split}
\end{equation}
\begin{equation}\label{psi_P1}
    \psi^{F}_{FM}(r,\theta,\phi)=R^{F}_F(r){\bf Y}^{F}_{FM}(\theta,\phi),
\end{equation}
\begin{equation}\label{psi_P0}
   \psi^{1}_{00}(r,\theta,\phi)=R^{1}_0(r){\bf Y}^{1}_{00}(\theta,\phi),
\end{equation}
where $R^{F-1}_{F}(r)$, $R^{F+1}_{F}(r)$, and $R^{F}_{F}(r)$
are the radial parts of the envelope wave functions. Furthermore,
\begin{equation}\label{vector_spherical_harmonics}
   {\bf Y}^L_{FM}(\theta,\phi)=\sum_{m_1,m_2}C^{F M}_{L m_1 1 m_2}Y_{Lm_1}(\theta,\phi)u_{m_2}
\end{equation}
are the vector spherical harmonics that can be expressed in terms of the usual
spherical harmonics $Y_{nm}(\theta,\phi)$ and the Clebsh-Gordon coefficients
$C^{j m}_{j_1 m_1 j_2 m_2}$ \cite{Varshalovich}. For the first two types of
functions \eqref{psi_S} and \eqref{psi_P1} only solutions with $F\geq 1$ are
possible. We will see that the functions \eqref{psi_S} are of a mixed type,
whereas the functions \eqref{psi_P1} and \eqref{psi_P0} are of a heavy and
light hole type, respectively.

For a formulation of the boundary conditions for the hole states
we take into account that the main contribution to the valence
band states in SiO$_2$ is given by $p$-orbitals
\cite{Chelikowsky1977} and for description of hole states outside
the Si nanocrystal we can use the same form of the Luttinger
Hamiltonian \eqref{Eq:Hamiltonian_valence}. It is known that the
hole masses at the valence band maximum in SiO$_2$ are pretty
large \cite{Chelikowsky1977}. For simplicity we choose them to be
equal to $m_v=5m_0$. Then the corresponding values of the
coefficients $A_o$ and $B_o$ of the Luttinger Hamiltonian
\eqref{Eq:Hamiltonian_valence} are $A_o=-\frac{1}{2m_v}$ and
$B_o=0$. In such a case we can formulate appropriate Bastard type
boundary conditions.

Inserting the functions $\psi^{F-1,F+1}_{FM}$ given by
Eq.~\eqref{psi_S} into the Schr\"{o}dinger equation with the
Hamiltonian \eqref{Eq:Hamiltonian_valence} we get the following
equation system (see also Ref.~\onlinecite{Baldereschi}) for the
radial functions $R^{F-1}_{F}(r)$, $R^{F+1}_{F}(r)$ inside the
nanocrystal ($r<R_{\rm nc}$):
\begin{widetext}
\begin{equation}\label{sys_RS}
  \begin{split}
    &\left(\!1+\frac{F\!-\!1}{2F\!+\!1}\mu\!\right)\left[\frac{\mbox{d}^2}{\mbox{d}r^2}
    +\frac{2}{r}\frac{\mbox{d}}{\mbox{d}r}-\frac{(F\!-\!1)F}{r^2}\right]R^{F-1}_{F}(r)\\
    &-\frac{3\sqrt{F(F\!+\!1)}}{2F\!+\!1}\mu\left[\frac{\mbox{d}^2}{\mbox{d}r^2}+\frac{2F\!+\!3}{r}\frac{\mbox{d}}{\mbox{d}r}
    +\frac{F(F\!+\!2)}{r^2}\right]R^{F+1}_{F}(r)
     \hspace{0cm}=-\frac{E}{A\hbar^2}R^{F-1}_{F}(r),
    \end{split}
\end{equation}
\begin{equation}\label{sys_RD}
    \begin{split}
    &-\frac{3\sqrt{F(F\!+\!1)}}{2F\!+\!1}\mu\left[\frac{\mbox{d}^2}{\mbox{d}r^2}
    -\frac{2F\!-\!1}{r}\frac{\mbox{d}}{\mbox{d}r}+\frac{(F\!-\!1)(F\!+\!1)}{r^2}\right]R^{F-1}_{F}(r)\\
    &+\left(\!1\!+\!\frac{F\!+\!2}{2F\!+\!1}\mu\!\right)\left[\frac{\mbox{d}^2}{\mbox{d}r^2}
    +\frac{2}{r}\frac{\mbox{d}}{\mbox{d}r}-\frac{(F\!+\!1)(F\!+\!2)}{r^2}\right]R^{F+1}_{F}(r)
    \hspace{0cm}=-\frac{E}{A\hbar^2}R^{F+1}_{F}(r),
    \end{split}
\end{equation}
\end{widetext}
where $E$ denote the hole energy and $\mu=B/A$. The general
solution of the equation system of Eqs. \eqref{sys_RS} and
\eqref{sys_RD}, which does not diverge at $r=0$, is found as
\begin{equation}\label{RS_sol}
 R^{F-1}_{F}(r)={\cal C}j_{F-1}(\lambda r/R_{\rm nc})+{\cal D}j_{F-1}(\lambda\beta r/R_{\rm nc}),
\end{equation}
\begin{equation}\label{RD_sol}
  \begin{split}
    R^{F+1}_{F}(r)=&-\sqrt{\frac{F}{F+1}}{\cal C} j_{F+1}(\lambda r/R_{\rm nc})\\
    &+\sqrt{\frac{F+1}{F}}{\cal D} j_{F+1}(\lambda\beta
    r/R_{\rm nc})\; ,
  \end{split}
\end{equation}
where ${\cal C}$, ${\cal D}$ are coefficients to be found from the
boundary and normalization conditions, $j_l(z)$ are the spherical
Bessel functions of the first kind \cite{Abramowitz},
and
\begin{equation}\label{Eq:beta}
    \beta=\sqrt{\frac{1-\mu}{1+2\mu}}\;.
\end{equation}
The
energy $E$, which is negative, is connected with the positive
variable $\lambda$ via
\begin{equation}\label{lambda}
    E=\frac{A\hbar^2}{R_{\rm nc}^2}(1-\mu)\lambda^2
    =-\frac{\hbar^2}{2m_hR_{\rm nc}^2}\lambda^2.
\end{equation}


Outside of the nanocrystal ($r>R_{\rm nc}$) the radial parts of
the functions $\psi^{F-1,F+1}_{FM}$ satisfy the following
equations
\begin{equation}\label{Eq:sys_RS_out}
   \begin{split}
    A_o\hbar^2\left[\frac{\mbox{d}^2}{\mbox{d}r^2}+\frac{2}{r}\frac{\mbox{d}}{\mbox{d}r}
    -\frac{(F-1)F}{r^2}\right]R^{F-1}_{F}(r)&\\
    =-(E+U_h)R^{F-1}_{F}(r)&\;,
    \end{split}
\end{equation}
\begin{equation}\label{Eq:sys_RD_out}
    \begin{split}
    A_o\hbar^2\left[\frac{\mbox{d}^2}{\mbox{d}r^2}
    +\frac{2}{r}\frac{\mbox{d}}{\mbox{d}r}-\frac{(F+1)(F+2)}{r^2}\right]R^{F+1}_{F}(r)&\\
    =-(E+U_h)R^{F+1}_{F}(r)&\; ,
    \end{split}
\end{equation}
where $U_h$ is the energy barrier for holes at the Si/SiO$_2$
boundary. According to Ref.~\onlinecite{Burdov} we have
$U_h=4.3$~eV. The general solution of the equation system of Eqs.
\eqref{Eq:sys_RS_out} and \eqref{Eq:sys_RD_out}, which converges
to zero for large distances from the nanocrystal is found as
\begin{equation}\label{Eq:RS_sol_out}
 R^{F-1}_{F}(r)=({\cal C}_o+{\cal D}_o) k_{F-1}(\kappa r/R_{\rm nc}),
\end{equation}
\begin{equation}\label{Eq:RD_sol_out}
    R^{F+1}_{F}(r)=\left(-\sqrt{\!\frac{F}{F\!+\!1}}\:{\cal C}_o+\sqrt{\frac{F\!+\!1}{F}}\:{\cal D}_0\right)k_{F+1}(\kappa r/R_{\rm nc}),
\end{equation}
where
\begin{equation}\label{Eq:kappa}
    \kappa=\frac{\sqrt{2m_v(E+U_h)}\;R_{\rm nc}}{\hbar},
    \end{equation}
${\cal C}_o$, ${\cal D}_o$ are again coefficients to be found from
the boundary and normalization conditions, and $k_l(z)$ are the
modified spherical Bessel functions of the third kind
\cite{Abramowitz}.
The boundary conditions lead to the following equations for the
radial functions:
\begin{widetext}
\begin{eqnarray}
   && R^{F-1}_{F}(r)\Big|_{r=R_{\rm
    nc}^-}=R^{F-1}_{F}(r)\Big|_{r=R_{\rm
    nc}^+}, \label{Eq:boundary_cond_SD1}\\
    && R^{F+1}_{F}(r)\Big|_{r=R_{\rm
    nc}^-}=R^{F+1}_{F}(r)\Big|_{r=R_{\rm
    nc}^+}, \label{Eq:boundary_cond_SD2}
\end{eqnarray}
\begin{equation}\label{Eq:boundary_cond_SD3}
    \left[\left(\!\left(\!A\!+\!\frac{F\!-\!1}{2F\!+\!1}B\!\right)
    \frac{\mbox{d}}{\mbox{d}r}+\frac{3}{2}\frac{F-1}{2F+1}\frac{B}{r}\!\right)R^{F-1}_{F}(r)
    -\frac{3\sqrt{F(F+1)}}{2F+1}B\left(\!\frac{\mbox{d}}{\mbox{d}r}+\frac{F+2}{r}\!\right)
        R^{F+1}_{F}(r)\right]_{r=R_{\rm nc}^-}\hspace{-0.3cm}=
    A_o\frac{\mbox{d}}{\mbox{d}r}R^{F-1}_{F}(r)\bigg|_{r=R_{\rm
    nc}^+},
\end{equation}
\begin{equation}\label{Eq:boundary_cond_SD4}
    \left[-\frac{3\sqrt{F(F+1)}}{2F+1}B\left(\!\frac{\mbox{d}}{\mbox{d}r}-\frac{F\!-\!1}{r}\!\right)
        R^{F-1}_{F}(r)
%
    +\left(\!\left(\!A\!+\!\frac{F\!+\!2}{2F\!+\!1}B\!\right)
    \frac{\mbox{d}}{\mbox{d}r}+\frac{3}{2}\frac{F+2}{2F+1}\frac{B}{r}\right)R^{F+1}_{F}(r)\right]_{r=R_{\rm nc}^-}
    \hspace{-0.5cm}=
    A_o\frac{\mbox{d}}{\mbox{d}r}R^{F+1}_{F}(r)\bigg|_{r=R_{\rm
    nc}^+}.
\end{equation}
\end{widetext}
Using the functions \eqref{RS_sol}, \eqref{RD_sol} and
\eqref{Eq:RS_sol_out}, \eqref{Eq:RD_sol_out} in Eqs.
\eqref{Eq:boundary_cond_SD1}-\eqref{Eq:boundary_cond_SD4} leads to
a solvability condition determining the energy eigenvalues for
states $\psi^{F-1,F+1}_{FM}$. We have derived this condition (see
Appendix). From it we have found numerically the energy
eigenvalues in dependence on the nanocrystal radius $R_{\rm nc}$.
The corresponding coefficients ${\cal C}, {\cal D}$, ${\cal C}_o$,
and ${\cal D}_o$, assuring the normalization condition, have been
also derived numerically.

\begin{figure}[t]
  \centering
  \epsfig{file=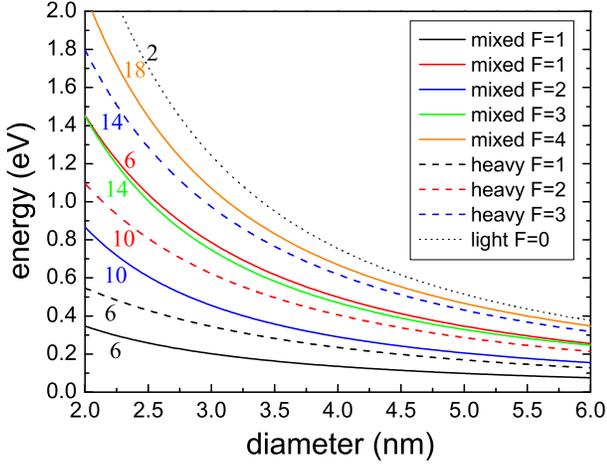,width=8.0cm}
  \caption{(Color online) Dependence of the positions of the
hole energy levels below the top of the valence band of bulk-Si on
the quantum dot diameter. The numbers near the lines indicate the
total degeneracy (including the spin degeneracy) of the
corresponding levels. Types of the levels and values of the total
angular momentum $F$ are also shown.}\label{Fig:levels_holes}
\end{figure}

For the radial function $R^{F}_{F}(r)$ of the states $\psi^{F}_{FM}$ we get the
following equation inside the Si quantum dot
\begin{equation}\label{Eq:RP_inside}
    (1-\mu)\left[\frac{\mbox{d}^2}{\mbox{d}r^2}+\frac{2}{r}\frac{\mbox{d}}{\mbox{d}r}
    -\frac{F(F+1)}{r^2}\right]R^{F}_{F}(r)
    =-\frac{E}{A\hbar^2}R^{F}_{F}(r).
\end{equation}
One can easily see that it is the same equation as for the radial
part of wave function of a particle having a simple parabolic band
with the heavy hole mass and angular momentum $F$. For the radial
function $R^{1}_{0}(r)$ of the states $\psi^{1}_{00}$ we get
\begin{equation}\label{Eq:RP0_inside}
   (1+2\mu)\left[\frac{\mbox{d}^2}{\mbox{d}r^2}+\frac{2}{r}\frac{\mbox{d}}{\mbox{d}r}-
   \frac{2}{r^2}\right]R^{1}_{0}(r)
    =-\frac{E}{A\hbar^2}R^{1}_{0}(r)\;.
\end{equation}
This equation is then the same as for a simple particle having the
light hole mass and angular momentum $1$. Solving equations
\eqref{Eq:RP_inside} and \eqref{Eq:RP0_inside} we find
\begin{equation}\label{Eq:RP1_sol}
 R^{F}_{F}(r)={\cal C}'j_F(\lambda r/R_{\rm nc}),
\end{equation}
\begin{equation}\label{Eq:RP0_sol}
 R^{1}_{0}(r)={\cal C}''j_{1}(\lambda \beta r/R_{\rm nc}),
\end{equation}
for $r<R_{\rm nc}$, where ${\cal C}'$ and ${\cal C}''$ are the
corresponding normalization coefficients. Outside the nanocrystal
$R^{F}_{F}(r)$ satisfies the equation
\begin{equation}\label{Eq:RP_outside}
    A_o\hbar^2\!\left[\frac{\mbox{d}^2}{\mbox{d}r^2}+\frac{2}{r}\frac{\mbox{d}}{\mbox{d}r}
    -\frac{F(F\!+\!1)}{r^2}\right]\!R^{F}_{F}(r)
    =-(E+U_h)R^{F}_{F}(r)
\end{equation}
and $R^{1}_{0}(r)$ satisfies the equation
\begin{equation}\label{Eq:RP0_outside}
    A_o\hbar^2\left[\frac{\mbox{d}^2}{\mbox{d}r^2}+\frac{2}{r}\frac{\mbox{d}}{\mbox{d}r}
    -\frac{2}{r^2}\right]R^{1}_{0}(r)
    =-(E+U_h)R^{1}_{0}(r).
\end{equation}
Solutions having an appropriate behavior at
infinity are
\begin{equation}\label{Eq:RP1_sol_outside}
 R^{F}_{F}(r)={\cal C}'_o k_F(\kappa r/R_{\rm nc}),
\end{equation}
\begin{equation}\label{Eq:RP0_sol_outside}
R^{1}_{0}(r)={\cal C}''_o k_{1}(\kappa r/R_{\rm nc}).
\end{equation}
The boundary conditions in this case are found as
\begin{equation}\label{Eq:boundary_cond_P1_1}
   R^{F}_{F}(r)\Big|_{r=R_{\rm
    nc}^-}=R^{F}_{F}(r)\Big|_{r=R_{\rm
    nc}^+},
\end{equation}
\begin{equation}\label{Eq:boundary_cond_P1_2}
    \left[(A-B)\frac{\mbox{d}}{\mbox{d}r}-\frac{3}{2}\frac{B}{r}\right]R^F_{F}(r)\bigg|_{r=R_{\rm
    nc}^-}\!\!\!=A_o\frac{\mbox{d}}{\mbox{d}r}R^{F}_{F}(r)\bigg|_{r=R_{\rm
    nc}^+},
\end{equation}
for $R^{F}_{F}(r)$ functions, and
\begin{equation}
   R^{1}_{0}(r)\Big|_{r=R_{\rm
    nc}^-}=R^{1}_{0}(r)\Big|_{r=R_{\rm
    nc}^+}, \label{Eq:boundary_cond_P0_1}
\end{equation}
\begin{equation}
    \left[(A+2B)\frac{\mbox{d}}{\mbox{d}r}+\frac{3B}{r}\right]R^{1}_{0}(r)\bigg|_{r=R_{\rm
    nc}^-}\!\!\!=A_o\frac{\mbox{d}}{\mbox{d}r}R^{1}_{0}(r)\bigg|_{r=R_{\rm
    nc}^+},\label{Eq:boundary_cond_P0_2}
\end{equation}
for $R^{1}_{0}(r)$ functions. Using the form of the radial wave
functions given by
Eqs.~\eqref{Eq:RP1_sol},\eqref{Eq:RP1_sol_outside}
[Eqs.~\eqref{Eq:RP0_sol},\eqref{Eq:RP0_sol_outside}] in
Eqs.~\eqref{Eq:boundary_cond_P1_1},\eqref{Eq:boundary_cond_P1_2}
[Eqs.~\eqref{Eq:boundary_cond_P0_1},\eqref{Eq:boundary_cond_P0_2}]
we get the equation determining the energy levels of states
$\psi^{F}_{FM}$ [$\psi^{1}_{00}$] (see Appendix). Solving this
equation numerically we have found energy level positions of holes
of the heavy and light hole types.

The dependence of positions of the hole energy levels on the
nanocrystal radius, their types and degeneracies are presented in
Fig.~\ref{Fig:levels_holes} for several lowest hole levels. One
can see that the hole level structure is more dense in comparison
with the electron level structure. This can lead to important
differences in behavior of ''hot'' electrons and holes
\cite{PRL_Tom}.


\section{Coulomb shift}
The Coulomb interaction leads to a decrease of the recombination energy of an
electron-hole pair \cite{Efros_exciton}. This interaction should be considered
taking into account the ''image charge'' effects appearing because of the
dielectric constant difference at the quantum dot boundary. On the other side
the dielectric constant mismatch leads to interaction of a charged particle
with its own image. Resulting polarization self-energy correction increases
single-particle energies of electrons and holes (counted upwards from the
bottom of the bulk conduction band and downwards from the top of the valence
band, respectively). If the dielectric constant changes discontinuously at the
quantum dot boundary the polarization self-interaction diverges there. This is
not an immediate problem for models of quantum dots assuming infinite energy
barriers for electrons and holes at the quantum dot boundary because particle
wave functions vanish there and the self-energy correction induced by the
self-interaction remains finite \cite{Brus,Lannoo1994,FTT}. However, as a
consequence of finite energy barriers particle wave functions are finite at the
quantum dot boundary (see Fig.~\ref{Fig:electron_functions}) and self-energy
corrections become infinite.

In order to remove these unphysical divergences one has to take
into account that the position-dependent dielectric constant
should change smoothly between the values corresponding to the
quantum dot core and the surrounding material on the scale of the
interatomic distance. This can be done in a simple and intuitive
way by ''regularizing'' directly the self-interaction
\cite{Koch_dielectric,Franceschetti_dielectric}.  A more physical
and controllable, however more demanding way is to solve the
problem with the smooth position-dependent dielectric constant
\cite{Bolcatto_dielectric}. Calculations of
Ref.~\onlinecite{Bolcatto_dielectric} show that the total
correction to the electron-hole pair energy introduced by the
Coulomb and ''image charge'' interactions depends strongly on the
width of the transition region between two values of the
dielectric constant. For reasonable values of the transition
region width around the interatomic distance the corrections
introduced by the electron-hole Coulomb interaction and the
polarization self-energy corrections cancel each other to high
extent. Therefore, the overall Coulomb correction to the energy of
the electron-hole pair is small and electron-hole recombination
energy is pretty accurately given by the sum of the band gap with
the electron and hole single-particle quantization energies. This
conclusion is also supported by GW calculations for small
hydrogen-terminated silicon nanocrystals \cite{Delerue_GW}.

An estimate (rather an upper limit estimate \cite{Bolcatto_dielectric}) of the
overall Coulomb shift $V_{\rm C}$ can be deduced using single-particle wave
functions corresponding to infinitely high energy barriers.
We have calculated $V_{\rm C}$ under these assumptions treating the Coulomb
correction as a perturbation to the single-particle Hamiltonian
\cite{FTT,EMRS2006_exciton}. The result for the ground-state electron and hole
can be given then in a simple form: $V_{\rm C}=-1.54{e^2}/(\kappa_{\rm Si}
R_{\rm nc})$, where $\kappa_{\rm Si}$ is the dielectric constant of Si and it
was taken into account that the surrounding material has an approximately three
times smaller dielectric constant than Si (one should notice that
Ref.~\onlinecite{FTT} gives a different numerical constant computed incorrectly
by us).
The excitonic shift calculated in the same way for higher states of electrons
and hole is on the same order and smaller.

\begin{figure}[t]
\centerline{\psfig{figure=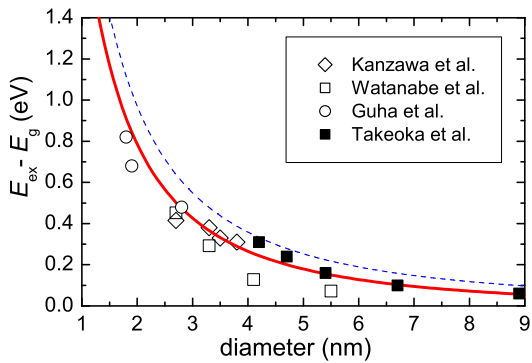,width=7.0cm}}
\caption{(Color online) Dependence of the ground-state electron-hole
recombination energy as a function of diameter of nanocrystal (solid line).
Dashed line shows the same energy without taking into account the exciton
shift. For comparison, the experimental data obtained from photoluminescence
spectra \cite{exp1,exp5,exp6,Watanabe2001} measured for Si nanocrystals inside
SiO$_2$ are presented.\label{Fig:exciton_energy}}
\end{figure}

In Fig.~\ref{Fig:exciton_energy} the dependence of the exciton
energy as a function of diameter of the nanocrystal is presented
for the ground state. This dependence  is also shown when
 the excitonic energy shift is accounted for. The
theory is compared with the experimental data obtained from the
photoluminescence spectra \cite{exp1,exp5,exp6,Watanabe2001}.

\begin{figure}[t]
\centerline{\psfig{figure=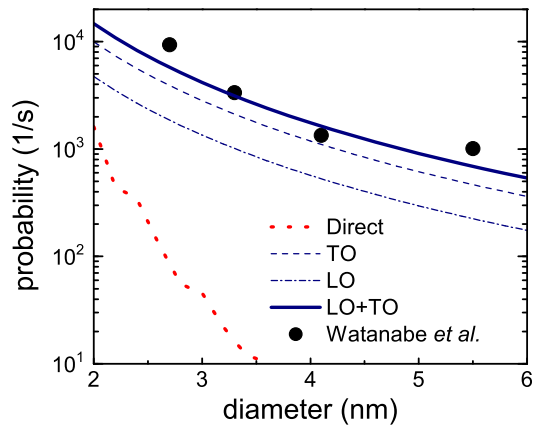,width=7.0cm}}
\caption{(Color online) Probabilities $P_{r,gr}$ of radiative transitions
between the ground electron and hole states: assisted by emission of a
TO-phonon (dashed line), an LO-phonon (dash-dot line) and their sum (thick
solid line) as well as probability of direct (zero-phonon) transition (dot
line), as functions of nanocrystal diameter. Experimental
points~\cite{Watanabe2001} are shown as
well.\label{Fig:transition_probabilities}}
\end{figure}

\section{Radiative recombination}
We have produced calculation of the probabilities $P_r$ of
radiative recombination assisted by emission of an optical
transverse phonon (with energy 57.5 meV) as well as longitudinal
one (55.3 meV). These channels of radiative transitions dominate
in bulk Si. The details of the calculation will be published in a
separate paper.
The results of calculations for the ground exciton state ($P_{r,gr}$) are
presented in Fig.~\ref{Fig:transition_probabilities}. The probabilities of
radiative transitions involving excited states $P_r$ have similar dependences
on nanocrystal size being of the same order of magnitude (e.g. for transition
from the second electron state to the first hole state $P_r \approx 0.8
P_{r,gr}$). In Fig.~\ref{Fig:transition_probabilities} the result of
calculations of direct (zero-phonon) radiative transition for the ground
exciton state is presented as well. Such a transition becomes possible for
confined carriers but one can see that the oscillator strength is noticeably
less for the dots with diameter larger than 2~nm. This is a well-known
experimental fact \cite{Kovalev1998}. One should notice that the shown
probability of the direct radiative transition has been calculated as an
average value over the nanocrystal size distribution in order to achieve the
acceptable convergence of the numerical integration and to avoid strong
oscillations of the result \cite{Hybertsen1994,Delerue_radiative}. One can see
from Fig.~\ref{Fig:transition_probabilities} that our results reproduce the
experimental data on radiative lifetimes in Si/SiO$_2$ nanocrystals
\cite{Watanabe2001} very well. As far as raditive transitions most probably
take place together with a phonon emission the exciton band gap derived from
 the photoluminescence spectra should generally be lower than the calculated one
 by the amount of the phonon
energy  (cf. Fig.~\ref{Fig:exciton_energy}).

\section{Conclusion}
We have calculated the wave functions and the energy levels of confined
carriers in Si quantum dots inside a SiO$_2$ matrix as functions of the dot
diameter. It has been shown that for small quantum dots ($d\lesssim$ 2.5~nm)
the energy spacing between neighboring electron and hole levels is of the order
of hundreds of meV,  and for electrons they are larger than for holes. Such
energy spacings are also larger than the energy level splittings due to
different mechanisms \cite{Niquet}, which are not accounted for in this paper.
Therefore, the single-phonon relaxation between the levels becomes impossible
and the time of relaxation of ''hot'' carriers to the ground state should
increase. The calculated recombination energies of an electron-hole pair in the
ground state and the probabilities of radiative interband transitions between
the ground electron and hole levels are in good agreement with experimental
data. Comparison of the electron and hole levels calculated by our method and
by state-of-the-art computational methods \cite{Garoufalis2006} could lead in
the future to a more rigorous formulation of the boundary conditions in the
framework of the multiband effective mass approximation. This would allow to
produce rather quantitative calculations of processes involving electrons and
holes in Si nanocrystals with a reasonable  computational effort.

\emph{Acknowledgement.-} We thank Prof. E.L. Ivchenko and Prof. V.I. Perel for
useful discussions. On the Russian side, this work was partly financially
supported by RFBR, NWO, INTAS, RAS. One of the authors (A.~A.~P.) was also
supported by the Dynasty Foundation.


\renewcommand{\theequation}{A\arabic{equation}}
\setcounter{equation}{0}  
\section*{APPENDIX: EQUATIONS FOR QUANTIZATION ENERGIES OF HOLES}

Equations determining the energy quantization levels for holes of mixed, heavy
and light holes are
\begin{widetext}
\begin{equation}\label{Eq:det_mixed}
\begin{split}
  &\left[ \nu\kappa \left( \frac {F-1}{2F-1}\frac{k_{{F-2}}(\kappa)}{k_{{F-1}}(\kappa)}
 +\frac
 {F}{2F-1}\frac{k_{{F}}(\kappa)}{k_{{F-1}}(\kappa)}\right)
j_{{F-1}} \left( \lambda \right) + \left( 1+\frac {F-1}{2F+1}\mu
\right) \lambda
\left(\frac{F-1}{2F-1}j_{{F-2}}(\lambda)-\frac {F}{2F-1}j_{{F}}(\lambda) \right)\right.\\
&\hspace{0.5cm}\left.+\mu\left(
\frac{3}{2}\frac{F-1}{2F+1}j_{{F-1}}(\lambda)
+\frac {3F(F+2)}{2F+1} j_{{F+1}}(\lambda) \right)
 +\frac{3F}{2F+1}
\mu\lambda \left(\frac {F\!+\!1}{2F\!+\!3}j_{{F}}(\lambda)
-\frac {F\!+\!2}{2F\!+\!3}j_{{F+2}}(\lambda) \right)\right]\\
& \times \left[  \frac{F\!+\!1}{F} \nu\kappa
\left(\frac{F\!+\!1}{2F\!+\!3}\frac{k_{{F}}(\kappa)}{k_{{F+1}}(\kappa)}
+\frac{F\!+\!2}{2F\!+\!3}\frac{k_{{F+2}}(\kappa)}{k_{{F+1}}(\kappa)}
\right)j_{{F+1}}(\lambda\beta)
-3\frac{F\!+\!1}{2F\!+\!1}\mu\lambda\beta\left(\frac
{F\!-\!1}{2F\!-\!1} j_{{F-2}}(\lambda\beta)
-\frac {F}{2F\!-\!1}j_{{F}}(\lambda\beta) \right) \right.\\
& \hspace{1.0cm} \left.
+\mu\left(\frac{3(F^2-1)}{2F+1}j_{{F-1}}(\lambda\beta)
+\frac{3}{2}\frac {(F+1)(F+2)}{F(2F+1)}j_{{F+1}}(\lambda\beta)
\right) \right.\\
& \hspace{1.0cm}\left.+ \frac{F+1}{F}  \left( 1+\frac
{F+2}{2F+1}\mu \right) \lambda\beta\left( \frac
{F+1}{2F+3}j_{{F}}(\lambda\beta)
-\frac{F+2}{2F+3}j_{{F+2}}(\lambda\beta)\right) \right]= \\
&\left[\nu\kappa\left(\frac{F-1}{2F-1}\frac{k_{{F-2}}(\kappa)}{k_{{F-1}}(\kappa)}
+\frac {F}{2F-1} \frac{k_{{F}}(\kappa)}{k_{{F-1}}(\kappa)}
\right)j_{{F-1}}(\lambda\beta) 
+ \left(1+\frac {F-1}{2F+1}\mu \right)\lambda\beta
\left(\frac{F-1}{2F-1}j_{{F-2}}(\lambda\beta)
-\frac {F}{2F-1}j_{{F}}(\lambda\beta) \right) \right.\\
&\hspace{0.5cm}\left.+\mu\left(
\frac{3}{2}\frac{F-1}{2F+1}j_{{F-1}}(\lambda\beta)
-\frac {3(F+1)(F+2)}{2F+1}j_{{F+1}}(\lambda\beta) \right)
-\frac{3(F+1)}{2F+1}\mu\lambda\beta\left(\frac{F+1}{2F+3}j_{{F}}(\lambda\beta)
-\frac {F+2}{2F+3}j_{{F+2}}(\lambda\beta)
 \right)   \right]\\
 &\times \left[ -\nu\kappa  \left(\frac {F+1}{2F+3} \frac{k_{{F}}(\kappa)}{k_{{F+1}}(\kappa)}
+\frac{F+2}{2F+3}\frac{k_{{F+2}}(\kappa)}{k_{{F+1}}(\kappa)}
\right)j_{{F+1}}(\lambda) 
-\frac{3(F+1)}{2F+1}\mu\lambda\left(\frac{F-1}{2F-1}j_{{F-2}}(\lambda)
-\frac {F }{2\,F-1} j_{{F}}(\lambda)
 \right)  \right.\\
&\hspace{1.0cm}\left.
+\mu\left(\frac{3(F^2-1)}{2F+1}j_{{F-1}}(\lambda)
-\frac{3}{2}\frac{F+2}{2F+1}j_{{F+1}}(\lambda)
\right)
-\left(1+\frac{F+2}{2F+1}\mu\right)\lambda
\left(\frac{F+1}{2F+3}j_{{F}}(\lambda)
-\frac{F+2}{2F+3}j_{{F+2}}(\lambda) \right)
 \right],
\end{split}
\end{equation}
\begin{equation}\label{Eq:det_heavy}
  \left(1-\mu\right) \lambda \left[ {\frac {F}{2F+1}}j_{{F-1}}(\lambda)
  -{\frac {F+1}{2F+1}}j_{{F+1}}(\lambda) \right] -\frac{3}{2}\mu j_{{F}} \left( \lambda
\right) +\nu\kappa j_{{F}}(\lambda) \left[ { \frac{F}{2F+1}
}\frac{k_{{F-1}}(\kappa)}{k_{{F}}(\kappa)} +
\frac{F+1}{2F+1}\frac{k_{{F+1}}(\kappa)}{k_{{F}} (\kappa)}
\right]=0,
\end{equation}
\begin{equation}\label{Eq:det_light}
  (1+2\mu)\lambda\beta\left[j_{{0}}(\lambda\beta)-2j_{{2}}(\lambda\beta)\right]k_{{1}}(\kappa)
  +9\mu j_{{1}}(\lambda\beta)k_{{1}}(\kappa)+\eta\kappa j_{{1}}(\lambda\beta)\left[k_{{0}}(\kappa)+
  2k_{{2}}(\kappa)\right]=0,
\end{equation}
\end{widetext}
respectively. Here $\mu=B/A$, $\nu=A_o/A$, $\beta$ is defined by
Eq.~\eqref{Eq:beta}, $\lambda$ and $\kappa$ are functions of the
hole energy $E$ determined by Eq.~\eqref{lambda} and
Eq.~\eqref{Eq:kappa}, respectively.


%

\end{document}